\documentclass{article}
\usepackage{ismir,amsmath,cite,url}
\usepackage{graphicx}
\usepackage{color}

\usepackage{booktabs}       
\usepackage{multirow}

\usepackage{marvosym}

\usepackage[dvipsnames]{xcolor}

\title{Learning Audio\ --\ Sheet Music Correspondences\\ for Score Identification and Offline Alignment}

\multauthor
{Matthias Dorfer$^*$ \hspace{1cm} Andreas Arzt$^{*\dagger}$ \hspace{1cm} Gerhard Widmer$^{*\dagger}$}
{$^*$Department of Computational Perception, Johannes Kepler University Linz, Austria\\
$^\dagger$The Austrian Research Institute for Artificial Intelligence (OFAI), Austria\\
{\tt matthias.dorfer@jku.at}
}
\def\authorname{Matthias Dorfer, Andreas Arzt, Gerhard Widmer}

\begin{document}

\maketitle
\begin{abstract}
This work addresses the problem of matching short excerpts of audio
with their respective counterparts in sheet music images.
We show how to employ neural network-based cross-modality embedding spaces
for solving the following two sheet music-related tasks:
retrieving the correct piece of sheet music from a database
when given a music audio as a search query; and
aligning an audio recording of a piece with the corresponding images of sheet music.
We demonstrate the feasibility of this in experiments on classical piano music by five different composers
(Bach, Haydn, Mozart, Beethoven and Chopin), and
additionally provide a discussion on
why we expect multi-modal neural networks to be a fruitful paradigm
for dealing with sheet music and audio at the same time.

\end{abstract}

\section{Introduction}
\label{sec:introduction}

Traditionally, automatic methods for linking audio and sheet music data are based on a common mid-level representation that allows for comparison (i.e., computation of distances or similarities) of time points in the audio and positions in the sheet music. Examples of mid-level representations are symbolic descriptions, which involve the error-prone steps of automatic music transcription on the audio side \cite{boeck:icassp:2012,kelz:ismir:2016,Sigtia2016Transcription,cheng2016attack} and optical music recognition (OMR) on the sheet music side \cite{wen2015omr,hajic:ismir:2016,byrd:jnmr:2015,rebelo:jmir:2012}, or spectral features like pitch class profiles (chroma features), which avoid the explicit audio transcription step but still depend on variants of OMR. For examples of the latter approach see, e.g., \cite{kurth2007automated,fremerey2009sheet,izmirli2012bridging}.

In this paper we present a methodology to \emph{directly} learn correspondences between complex audio data and images of the sheet music, circumventing the problematic definition of a mid-level representation. Given short snippets of audio and their respective sheet music images, a cross-modal neural network is trained to learn an embedding space in which both modalities are represented as 32-dimensional vectors. which can then be compared, e.g., via their cosine distance. Essentially, the neural network replaces the complete feature computation process (on both sides) by learning a transformation of data from the audio and from the sheet music to a common vector space.

The idea of matching sheet music and audio with neural networks was recently proposed in \cite{Dorfer2016Towards}. The approach presented here goes beyond that in several respects.
First, the network in \cite{Dorfer2016Towards} requires both sheet music and audio as input at the same time
to predict which location in the sheet image best matches the current audio excerpt. We address a more general scenario where both input modalities are required only at training time, for learning the relation between score and audio. This requires a different network architecture that can learn two separate projections, one for embedding the sheet music and one for embedding the audio.
These can then be used independently of each other.
For example, we can first embed a reference collection of sheet music images using the image embedding part of the network, then embed a query audio and search for its nearest sheet music neighbours in the joint embedding space.
This general scenario is referred to as \emph{cross-modality retrieval} and supports different applications (two of which are demonstrated in this paper).
The second aspect in which we go beyond \cite{Dorfer2016Towards} is the sheer complexity of the musical material: while \cite{Dorfer2016Towards} was restricted to simple monophonic melodies, we will demonstrate the power of our method on real, complex pieces of classical music.

We demonstrate the utility of our approach via preliminary results on two real-world tasks. The first is \emph{piece identification:} given an audio rendering of a piece, the corresponding sheet music is identified via cross-modal retrieval. (We should note here that for practical reasons, in our experiments the audio data is synthesized from MIDI -- see below). The second task is \emph{audio-to-sheet-music alignment}. Here, the trained network acts as a complex distance function for given pairs of audio and sheet music snippets, which in turn is used by a dynamic time warping algorithm to compute an optimal sequence alignment.









\begin{figure*}[ht!]
 \centerline{\includegraphics[width=1.8\columnwidth]{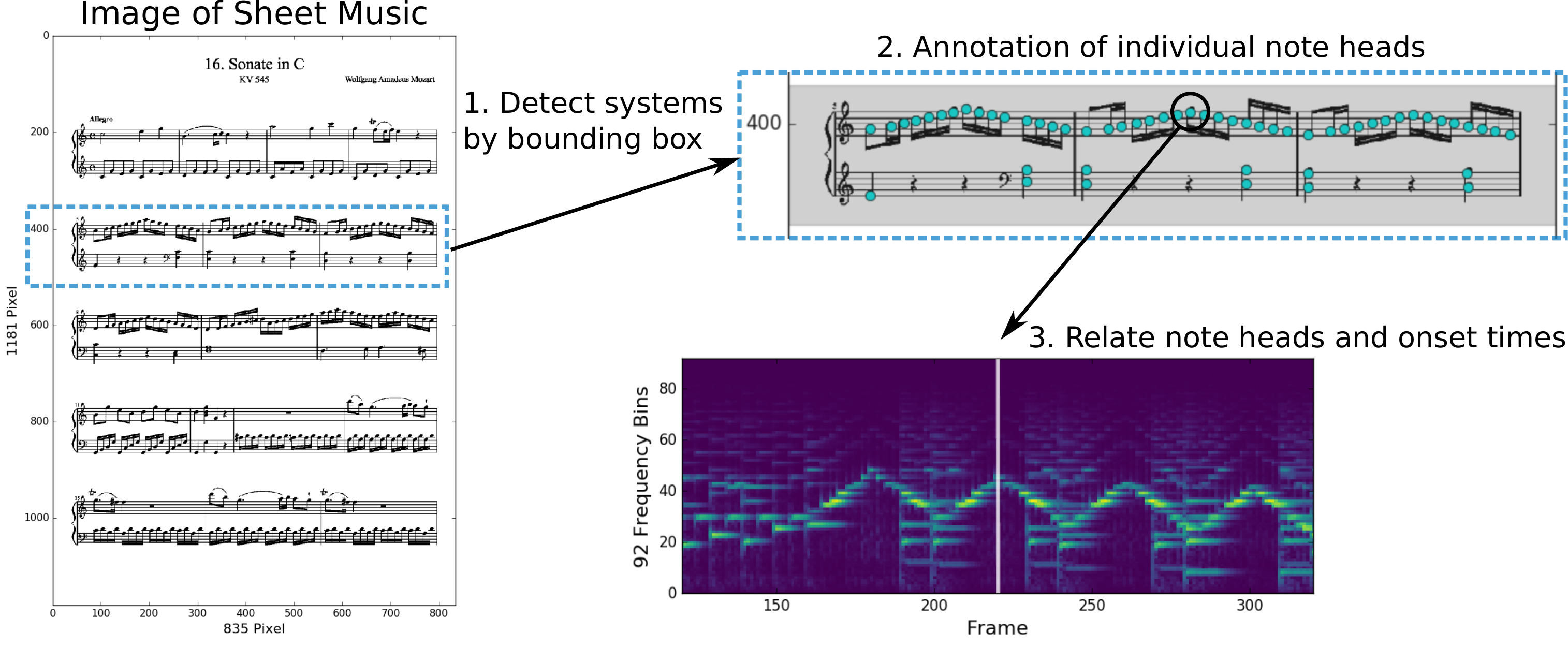}}
 \caption{Work flow for preparing the training data (correspondences between sheet music images and the respective music audio).
          Given the relation between the note heads in the sheet music image
          and their corresponding onset times in the audio signal
          we sample audio-sheet-music pairs for training our networks.
          Figure \ref{fig:corr_examples} shows four examples of such training pairs.}
\label{fig:a2s_correspondence}
\end{figure*}
Our main contributions, then, are (1) a methodology for learning cross-modal embedding spaces for relating audio data and sheet music data; (2) data augmentation strategies which allow for training the neural network for this complex task even with a limited amount of data; and (3) first results on two important MIR tasks, using this new approach.

\section{Description of Data}
\label{sec:data}

Our approach is built around a neural network designed for learning the relationship between two different data modalities.
The network learns its behaviour
solely from the examples shown for training.
As the presented data is crucial to make this class of models work,
we dedicate this section to describing the underlying data as well as the necessary preparation steps needed to generate training examples for optimizing our networks.

%
\subsection{Sheet-Music-Audio Annotation}
\label{subsec:annotation}
As already mentioned, we want to address two tasks:
(1) sheet music (piece) identification from audio queries and
(2) offline alignment of a given audio with its corresponding sheet music image.
Both are multi-modal problems involving sheet music images and audio.
We therefore start by describing the process of producing the ground truth for learning correspondences between a given score and its respective audio.
Figure \ref{fig:a2s_correspondence} summarizes the process.

Step one is the localization of staff systems in the sheet music images.
In particular, we annotate bounding boxes around the individual systems.
Given the bounding boxes we detect the positions of the note heads
within each of the systems\footnote{We of course do not annotate all of the systems and note heads by hand but use a note head and staff detector to support this tasks (again a neural network trained for this purpose).}.
The next step is then to relate the note heads to their corresponding onset times in the audio.

\begin{figure}[t!]
 \centerline{\includegraphics[width=1.0\columnwidth]{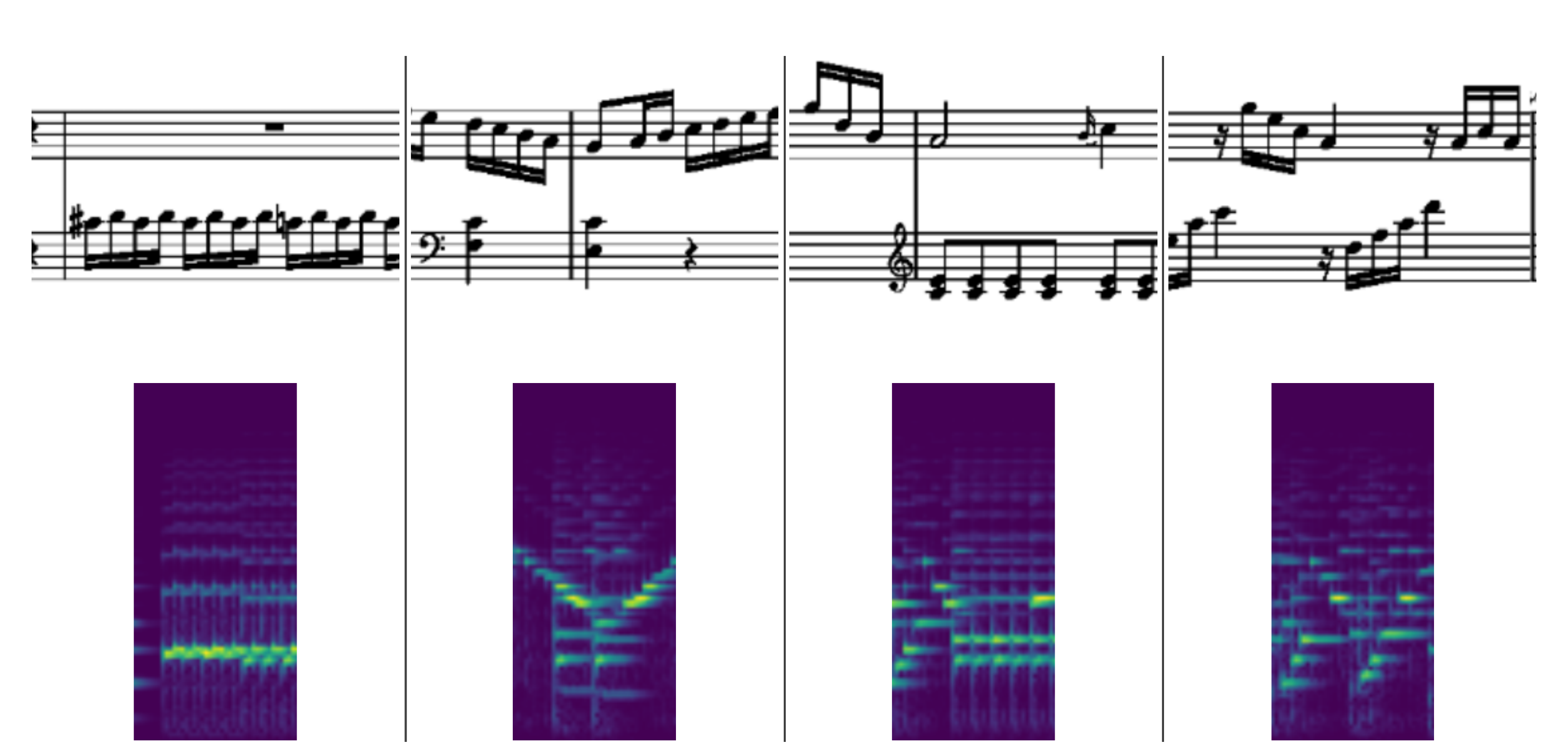}}
 \caption{Sheet-music audio correspondences presented to the network for retrieval embedding space learning.}
\label{fig:corr_examples}
\end{figure}

Once these relations are established, we know for each note head its location (in pixel coordinates) in the image,
and its onset time in the audio.
Based on this relationship we cut out corresponding snippets of sheet music images (in our case $180 \times 200$ pixels)
and short excerpts of audio represented by log-frequency spectrograms ($92$ bins $\times$ $42$ frames).
Figure \ref{fig:corr_examples} shows four examples of such sheet-music-audio correspondences;
these are the pairs presented to our multi-modal networks for training.
%

%
\subsection{Composers, Sheet Music and Audio}
\label{subsec:data}
For our experiments we use classical piano music by five different composers: Mozart (14 pieces), Bach (16), Beethoven (5), Haydn (4) and Chopin (1).
%
%
To give an impression of the complexity of the music, we have, for instance,
Mozart piano sonatas (K.545 1st mvt., K.331 3rd) and symphony transcriptions for piano (Symphony 40 K.550 1st), preludes and fugues from Bach's WTC,
Beethoven piano sonata movements
and Chopin's Nocturne Op.9 No.1.
In terms of experimental setup we will use \emph{only} the 13 pieces of Mozart for training,
Mozart's K.545 mvt.1 for validation,
and all remaining pieces for testing.
This results in 18,432 correspondences for training, 989 for validating,
and 11,821 for testing.
Our sheet music is collected from \emph{Musescore}\footnote{\url{https://musescore.com}} where we
selected only scores having a `realistic' layout close to the typesetting of professional publishers\footnote{This is an example of a typical score we used for the experiment (Beethoven Sonata Op.2 No.1): \url{https://musescore.com/classicman/scores/55331}}.
The reason for using Musescore for initial experiments is that along with the sheet music (as \emph{pdf} or image files) Musescore
also provides the corresponding \emph{midi} files.
This allows us to synthesize the music for each piece of sheet music
and to compute the exact note onset times from the midis, and thus
to establish the required sheet-music audio correspondences.

In terms of audio preparation we compute log-frequency spectrograms of the audios, with a sample rate of $22.05$kHz, a FFT window size of $2048$ samples, and a computation rate of 20 frames per second.
For dimensionality reduction we apply a normalized 16-band logarithmic filterbank allowing only frequencies from $30Hz$ to $16kHz$,
which results in 92 frequency bins.

%
\subsection{Data Augmentation}
\label{subsec:data_augmentation}
To improve the generalization ability of the resulting networks, we propose several data augmentation strategies specialized to score images and audio.
In machine learning, \emph{data augmentation} refers to the application of (realistic) data transformations in order to synthetically increase the effective size of the training set.
We already emphasize at this point that data augmentation is a crucial component for learning cross-modality representations that generalize to unseen music, especially when little data is available.

\begin{figure}[ht!]
 \centerline{\includegraphics[width=0.9\columnwidth]{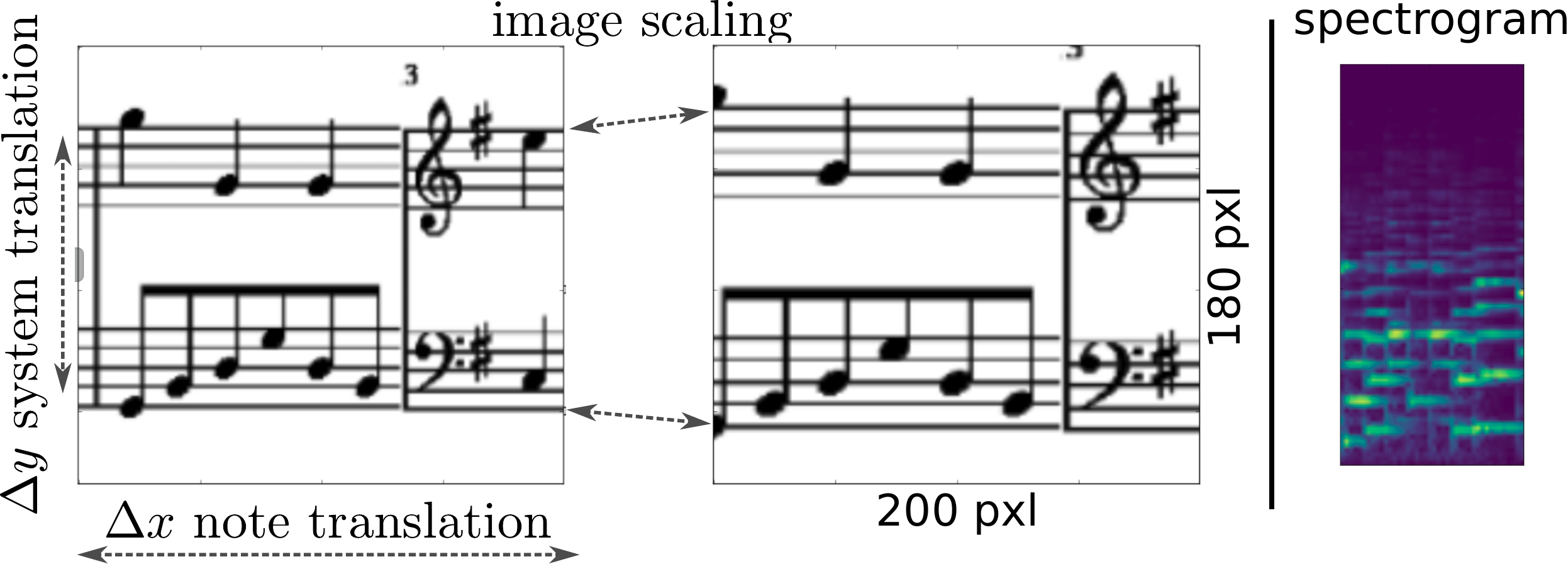}}
 \caption{Overview of image augmentation strategies.
          The size of the sliding image window remains constant ($180 \times 200$ pixels)
          but its content changes depending on the augmentations applied.
          The spectrogram remains the same for the augmented image versions.}
\label{fig:image_augmentation}
\end{figure}

For \textbf{sheet image augmentation} we apply three different transformations, summarized in Figure \ref{fig:image_augmentation}.
The first is \emph{image scaling} where we resize the image between $95$ and $105\%$ of its original size.
This should make the model robust to changes in the overall dimension of the scores.
Secondly, in \emph{$\Delta y$ system translation}
we slightly shift the system in the vertical direction by $\Delta y \in [-5, 5]$ pixels.
We do this as the system detector will not detect each system in exactly the same way
and we want our model to be invariant to such translations.
In particular, it should not be the absolute location of a note head in the image that determines its meaning (pitch) but its relative position with respect to the staff.
Finally, we apply \emph{$\Delta x$ note translation}, meaning that we slightly shift
the corresponding sheet image window by $\Delta x \in [-5, 5]$ pixels in the horizontal direction.

In terms of \textbf{audio augmentation} we render the training pieces with three different sound fonts
and additionally vary the tempo between 100 and 130 beats per minute (bpm).
The test pieces are all rendered at a rate of 120 bpm using an \emph{additional unseen sound font}.
The test set is kept fixed to reveal the impact of the different data augmentation strategies.

\section{Audio - Sheet Music Correspondence Learning}
\label{sec:methods}
This section describes the underlying learning methodology.
As mentioned above, the core of our approach is a cross-modality retrieval neural network capable of learning relations between short snippets of audio and sheet music images.
In particular, we aim at learning a joint embedding space of the two modalities in which to perform nearest-neighbour search.
One method for learning such a space,
which has already proven to be effective in other domains such as text-to-image retrieval, is based on the optimization of a pairwise ranking loss \cite{kiros2014unifying, socher2014grounded}.
Before explaining this optimization target,
we first introduce the general architecture of our correspondence learning network.
\begin{figure}[ht!]
 \centerline{\includegraphics[width=0.70\columnwidth]{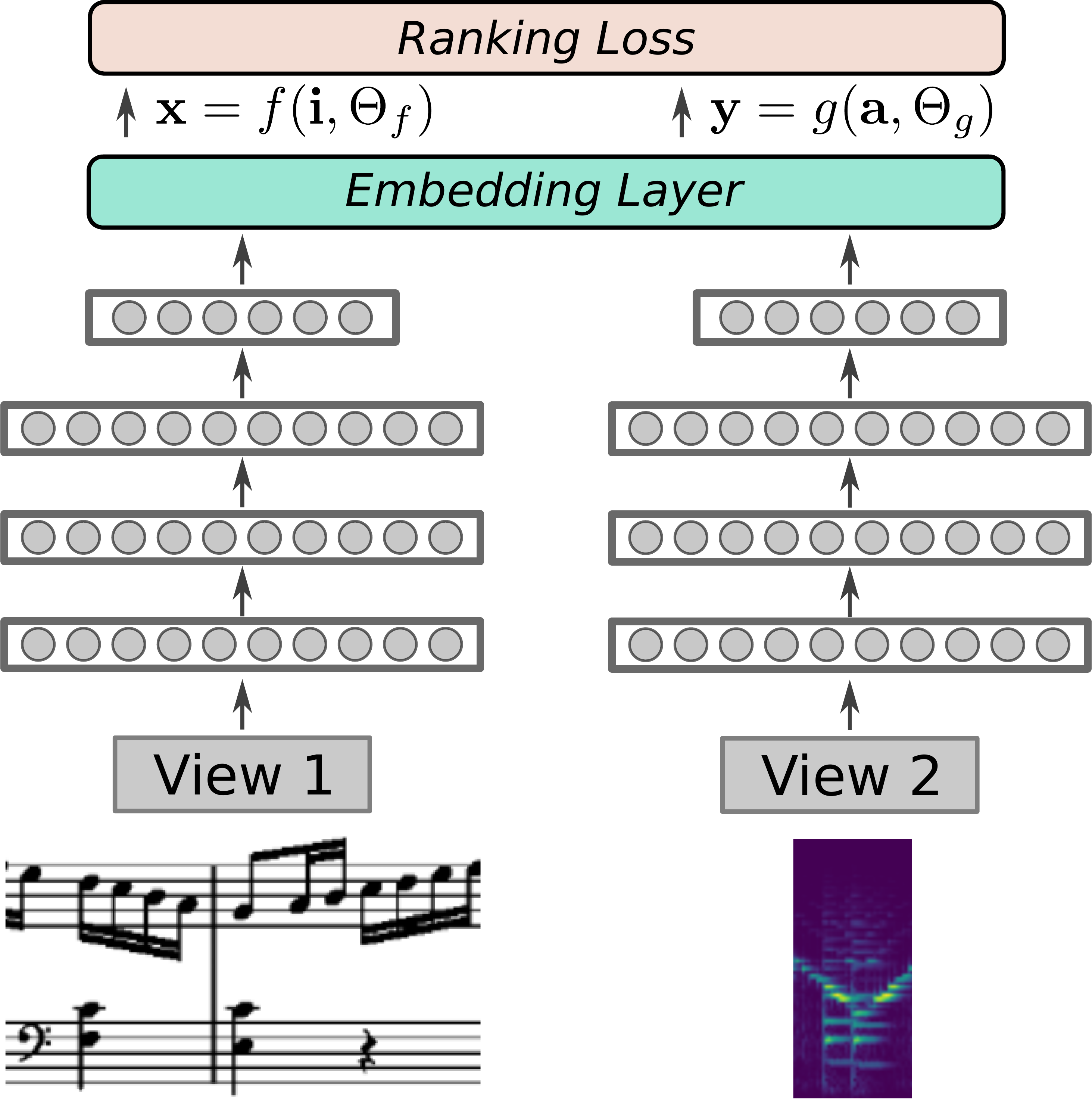}}
 \caption{Architecture of correspondence learning network.
          The network is trained to optimize the similarity (in embedding space) between corresponding
          audio and sheet image snippets by minimizing a pair-wise ranking loss.
}
\label{fig:model_architecture}
\end{figure}

As shown in Figure \ref{fig:model_architecture} the network consists of two separate pathways $f$ and $g$ taking two inputs at the same time. Input one is a sheet image snippet $\mathbf{i}$ and input two is an audio excerpt $\mathbf{a}$.
This means in particular that network $f$ is responsible for processing the image part of an input pair and network $g$ is responsible for processing the audio.
The output of both networks (represented by the \emph{Embedding Layer} in Figure \ref{fig:model_architecture}) is a $k$-dimensional vector representation encoding the respective inputs.
In our case the dimensionality of this representation is 32.
We denote these hidden representations by $\mathbf{x} = f(\mathbf{i, \Theta_f})$ for the sheet image
and $\mathbf{y} = g(\mathbf{a, \Theta_g})$ for the audio spectrogram, respectively,
where $\Theta_f$ and  $\Theta_g$ are the parameters of the two networks.

Given this network design, we now explain the pairwise ranking objective.
Following \cite{kiros2014unifying} we first introduce a \emph{scoring function} $s(\mathbf{x}, \mathbf{y})$
as the cosine similarity $\mathbf{x} \cdot \mathbf{y}$ between the two hidden representations
($\mathbf{x}$ and $\mathbf{y}$ are scaled to have unit norm).
Based on this scoring function we optimize the following pairwise ranking objective (`hinge loss'):
\begin{equation}
\mathcal{L}_{rank}=\sum_{\mathbf{x}} \sum_{k} \max \{0, \alpha - s(\mathbf{x}, \mathbf{y}) + s(\mathbf{x}, \mathbf{y}_k) \}
\label{eq:contrastive}
\end{equation}
In our application $\mathbf{x}$ is an embedded sample of a sheet image snippet,
$\mathbf{y}$ is the embedding of the matching audio excerpt
and $\mathbf{y}_k$ are the embeddings of the \emph{contrastive} (mismatching) audio excerpts
(in practice all remaining samples of the current training batch).
The intuition behind this loss function is to encourage an embedding space
where the distance between matching samples is lower than the distance between mismatching samples.
If this condition is roughly satisfied,
we can then perform cross-modality retrieval by simple nearest neighbour search in the embedding space.
This will be explained in detail in Section \ref{sec:eval_methods}.

The network itself is implemented as a VGG- style convolution network \cite{simonyan2014very} consisting of $3 \times 3$ convolutions followed by $2 \times 2$ max-pooling
as outlined in detail in Table \ref{tab:model_architecture}.
The final convolution layer computes 32 feature maps and is subsequently processed with a global average pooling layer \cite{LinCY2013NIN} that produces a 32-dimensional vector for each input image and spectrogram, respectively.
This is exactly the dimension of our retrieval embedding space.
At the top of the network we put a canonically correlated embedding layer \cite{dorfer2017end} combined with the ranking loss described above.
In terms of optimization we use the \emph{adam} update rule \cite{Kingma2014adam} with an initial learning rate of $0.002$.
We watch the performance of the network on the validation set and halve the learning rate if there is no improvement for 30 epochs.
This procedure is repeated ten times to finetune the model.

\begin{table}[ht]
\small
\caption{\small Audio-sheet-music model.
BN: Batch Normalization \cite{LoffeS2015BatchNorm}, ELU: Exponential Linear Unit \cite{clevert2015ELU},
MP: Max Pooling, Conv($3$, pad-1)-$16$: $3 \times 3$ convolution, 16 feature maps and padding 1.}
\vspace*{2mm}
\scriptsize
\centering
\begin{tabular}{c|c}
\hline
Sheet-Image $180 \times 200$ & Audio (Spectrogram) $92 \times 42$ \\
\hline
$2\times$Conv($3$, pad-1)-$12$	&	$2\times$ Conv($3$, pad-1)-$12$ \\
BN-ELU + MP($2$)					&	BN-ELU + MP($2$) \\
$2\times$Conv($3$, pad-1)-$24$	&	$2\times$ Conv($3$, pad-1)-$24$ \\
BN-ELU + MP($2$)					&	BN-ELU + MP($2$) \\
$2\times$Conv($3$, pad-1)-$48$	&	$2\times$ Conv($3$, pad-1)-$48$ \\
BN-ELU + MP($2$)					&	BN-ELU + MP($2$) \\
$2\times$Conv($3$, pad-1)-$48$	&	$2\times$ Conv($3$, pad-1)-$48$ \\
BN-ELU + MP($2$)					&	BN-ELU + MP($2$) \\
Conv($1$, pad-0)-$32$-BN-LINEAR	& Conv($1$, pad-0)-$32$-BN-LINEAR \\
GlobalAveragePooling 					& GlobalAveragePooling \\
\hline
\multicolumn{2}{c}{Embedding Layer + Ranking Loss} \\
\end{tabular}
\label{tab:model_architecture}
\end{table}

\section{Evaluation of Audio - Sheet Correspondence Learning}
\label{sec:eval_methods}
\begin{figure}[t!]
 \centerline{\includegraphics[width=0.9\columnwidth]{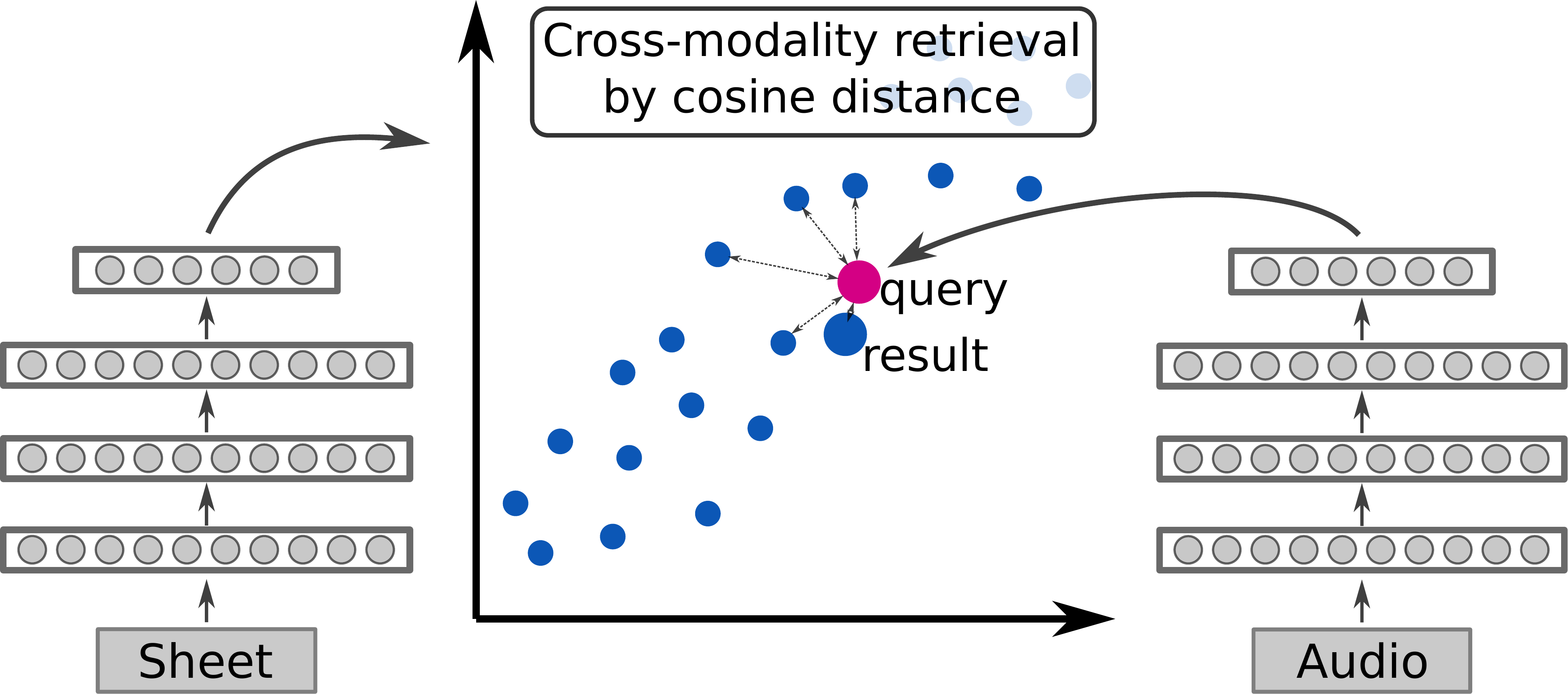}}
 \caption{Sketch of sheet-music-from-audio retrieval.
          The blue dots represent the embedded candidate sheet music snippets.
          The red dot is the embedding of an audio query.
          The larger blue dot highlights the closest sheet music snippet candidate
          selected as retrieval result.}
\label{fig:retrieval}
\end{figure}

In this section we evaluate the ability of our model to retrieve the correct counterpart
when given an instance of the other modality as a search query.
This first set of experiments is carried out on the lowest possible granularity, namely, on sheet image snippets and spectrogram excerpts such as shown in Figure \ref{fig:corr_examples}.
For easier explanation we describe the retrieval procedure from an \emph{audio query point of view}
but stress that the opposite direction works in exactly the same fashion.
Given a spectrogram excerpt $\mathbf{a}$ as a search query we want to retrieve
the corresponding sheet image snippet $\mathbf{i}$.
For retrieval preparation we first embed all candidate image snippets $\mathbf{i}_j$
by computing $\mathbf{x}_j=f(\mathbf{i}_j)$ as the output of the image network.
In the present case, these candidate snippets originate from the 26 unseen test pieces by Bach, Haydn, Beethoven and Chopin.
In a second step we embed the given query audio as $\mathbf{y}=g(\mathbf{a})$ using the audio pathway $g$ of the network.
Finally, we select the audio's nearest neighbour $\mathbf{x}_j$ from the set of embedded image snippets as
\begin{equation}
\mathbf{x}_j = \underset{\mathbf{x}_i}{\arg \min} \left(1.0 - \frac{\mathbf{x}_i \cdot \mathbf{y}}{\lvert\lvert \mathbf{x}_i \rvert\rvert \, \lvert\lvert \mathbf{y} \rvert\rvert }\right)
\label{eq:cos_dist}
\end{equation}
based on their pairwise cosine distance.
Figure \ref{fig:retrieval} shows a sketch of this retrieval procedure.

In terms of experimental setup we use the 13 pieces of Mozart for training
	the network, and the pieces of the remaining composers for testing.
As evaluation measures we compute the \emph{Recall@k (R@k)} as well as the \emph{Median Rank (MR)}.
The \emph{R@k} rate (high is better) is the percentage of queries which have the correct corresponding counterpart in the first $k$ retrieval results.
The \emph{MR} (low is better) is the median position of the target in a cosine-similarity-ordered list of available candidates.

Table \ref{tab:eval_methods} summarizes the results
for the different data augmentation strategies described in Section \ref{subsec:data_augmentation}.
The unseen synthesizer and the tempo for the test set remain fixed for all settings. This allows us to directly investigate the influence of the different augmentation strategies.
\begin{table}[t]
 \begin{center}
 \begin{tabular}{rcccc}
 \toprule
 \textbf{Audio Augmentation} & \textbf{R@1} & \textbf{R@10} & \textbf{R@25} & \textbf{MR} \\
 \midrule
 1 Synth, 100-130bpm 	& 0.37  & 3.73 & 7.05 & 771 \\
 3 Synth, 120bpm 		& 0.75	& 6.26 & 11.52 & 559 \\
 3 Synth, 100-130bpm 	& 0.87  & 8.23 & 15.29 & 332 \\
  \\
 \toprule
 \textbf{Sheet Augmentation} & & & & \\
 \midrule
 image scaling					& 0.75 & 5.60 & 10.14 & 524 \\
 $\Delta y$ system translation	& 0.91 & 6.57 & 12.21 & 449 \\
 $\Delta x$ note translation	& 0.44 & 3.66 & 7.19 & 808 \\
 full sheet augmentation		& 0.70 & 5.72 & 11.03 & 496 \\
 &&&& \\
 \toprule
 no augmentation				& 0.33 & 2.88 & 5.71 & 1042 \\
 full augmentation				& 1.70 & 11.67 & 21.17 & 168 \\
 \toprule
 random baseline				& 0.00 & 0.03 & 0.21 & 5923 \\
 \bottomrule 
 \end{tabular}
\end{center}
 \caption{Influence of data augmentation on audio-to-sheet retrieval.
          For the audio augmentation experiments no sheet augmentation is applied and vice versa.
          \emph{no augmentation} represents  1 Synth, 120bpm without sheet augmentation.}
 \label{tab:eval_methods}
\end{table}
The results are grouped into audio augmentation, sheet augmentation, and applying all or no data augmentation at all.
On first sight the retrieval performance appears to be very poor.
In particular the \emph{MR} seems hopelessly high in view of our target applications.
However, we must remember that our query length is only 42 spectrogram frames ($\approx$ 2 seconds of audio) per excerpt and
we select from a set of $11,821$ available candidate snippets.
And we will see in the following sections that this retrieval performance is still sufficient to perform tasks such as piece identification.
Taking the performance of \emph{no augmentation} as a baseline we observe that all data augmentation strategies help improve the retrieval performance.
In terms of audio augmentation we observe that training the model with different synthesizers and varying the tempo
works best.
From the set of image augmentations, the \emph{$\Delta y$ system translation} has the highest impact on retrieval performance.
Overall we get the best retrieval model when applying \emph{all} augmentation strategies.
Note also the large gap between \emph{no augmentation} and \emph{full augmentation}.
The median rank, for example, drops from 1042 in case of no augmentation to 168 for full augmentation, which is a substantial improvement.

A final note: for space reasons we only present results on audio-to-sheet music retrieval, but that the opposite direction using image snippets as search query works analogously and shows similar performance.

\section{Piece Identification}
\label{sec:retrieval}
Given the above model that learns to express similarities between sheet music snippets and audio excerpts,
we now describe how to use this to solve our first targeted task:
identifying the respective piece of sheet music when given an entire audio recording as a query (despite the relatively poor recall and MR for individual queries).

\subsection{Description of Approach}
We start by preparing a \textbf{sheet music retrieval database} as follows.
Given a set of sheet music images along with their annotated systems we cut each piece of sheet music $j$
into a set of image snippets $\{ \mathbf{i}_{ji} \}$ analogously to the snippets presented to our network for training.
For each snippet we store its originating piece $j$.
We then embed all candidate image snippets into the retrieval embedding space
by passing them through the image part $f$ of the multi-modal network.
This yields, for each image snippet, a 32-dimensional embedding coordinate vector $\mathbf{x}_{ji}=f(\mathbf{i}_{ji})$.

\textbf{Sheet snippet retrieval from audio:} Given a whole audio recording as a search query
we aim at identifying the corresponding piece of sheet music in our database.
As with the sheet image we start by cutting the audio (spectrogram) into a set of excerpts $\{ \mathbf{a}_1, ..., \mathbf{a}_K \}$
again exhibiting the same dimensions as the spectrograms used for training,
and embed all query spectrogram excerpts $\mathbf{a}_k$
with the audio network $g$.
Then we proceed as described in Section \ref{sec:eval_methods} and
select for each audio its nearest neighbour from the set of all embedded image snippets.

\textbf{Piece selection:} Since we know for each of the image snippets
its originating piece $j$, we can now have the retrieval image snippets $\mathbf{x}_{ji}$ \emph{vote} for the piece.
The piece achieving the highest count of votes is our final retrieval result.
In our experiments we consider for each query excerpt its top 25 retrieval results for piece voting.

\subsection{Evaluation of Approach}
Table \ref{tab:piece_retrieval} summarizes the piece identification results on our test set of Bach, Haydn, Beethoven and Chopin (26 pieces).
Again, we investigate the influence of data augmentation and observe that the trend of the experiments in Section \ref{sec:eval_methods} is directly reflected in the piece retrieval results.
As evaluation measure we compute $Rk$ as the number of pieces ranked at position $k$
when sorting the result list by the number of votes.
Without data augmentation only four of the 26 pieces are ranked first in the retrieval lists
of the respective full audio recording queries.
When making use of data augmentation during training, this number increases substantially
and we are able to recognize 24 pieces at position one; the remaining two are ranked at position two.
Although this is not the most sophisticated way of employing our network for piece retrieval, it clearly shows the usefulness of our model and its learned audio and sheet music representations for such tasks.

\begin{table}[t]
 \begin{center}
 \begin{tabular}{rcccc}
 \toprule
 \textbf{Augmentation} & \textbf{R1} & \textbf{R2} & \textbf{R3} & \textbf{$>$R3} \\
 \midrule
 no augmentation 			& 4  & 7 & 1 & 14 \\
 full augmentation 			& 24 & 2 & 0 & 0  \\
 \bottomrule
 \end{tabular}
\end{center}
 \caption{Influence of data augmentation on piece retrieval.}
 \label{tab:piece_retrieval}
\end{table}

\section{Audio-to-Sheet-Music Alignment}
\label{sec:alignment}
As a second usage scenario for our approach we present the task of audio-to-sheet-music alignment. Here, the goal is to align a performance (given as an audio file) to its respective score (as images of the sheet music), i.e., computing the corresponding location in the sheet music for each time point in the performance, and vice versa.

\subsection{Description of Approach}
For computing the actual alignments we rely on Dynamic Time Warping (DTW), which is a standard method for sequence alignment \cite{rabiner:book:1993}, and is routinely used in the context of music processing \cite{muller:book:2015}. Generally, DTW takes two sequences as input and computes an optimal non-linear alignment between them, with the help of a local cost measure that relates points of the two sequences to each other.

\begin{figure}[t!]
 \centerline{\includegraphics[width=1.0\columnwidth]{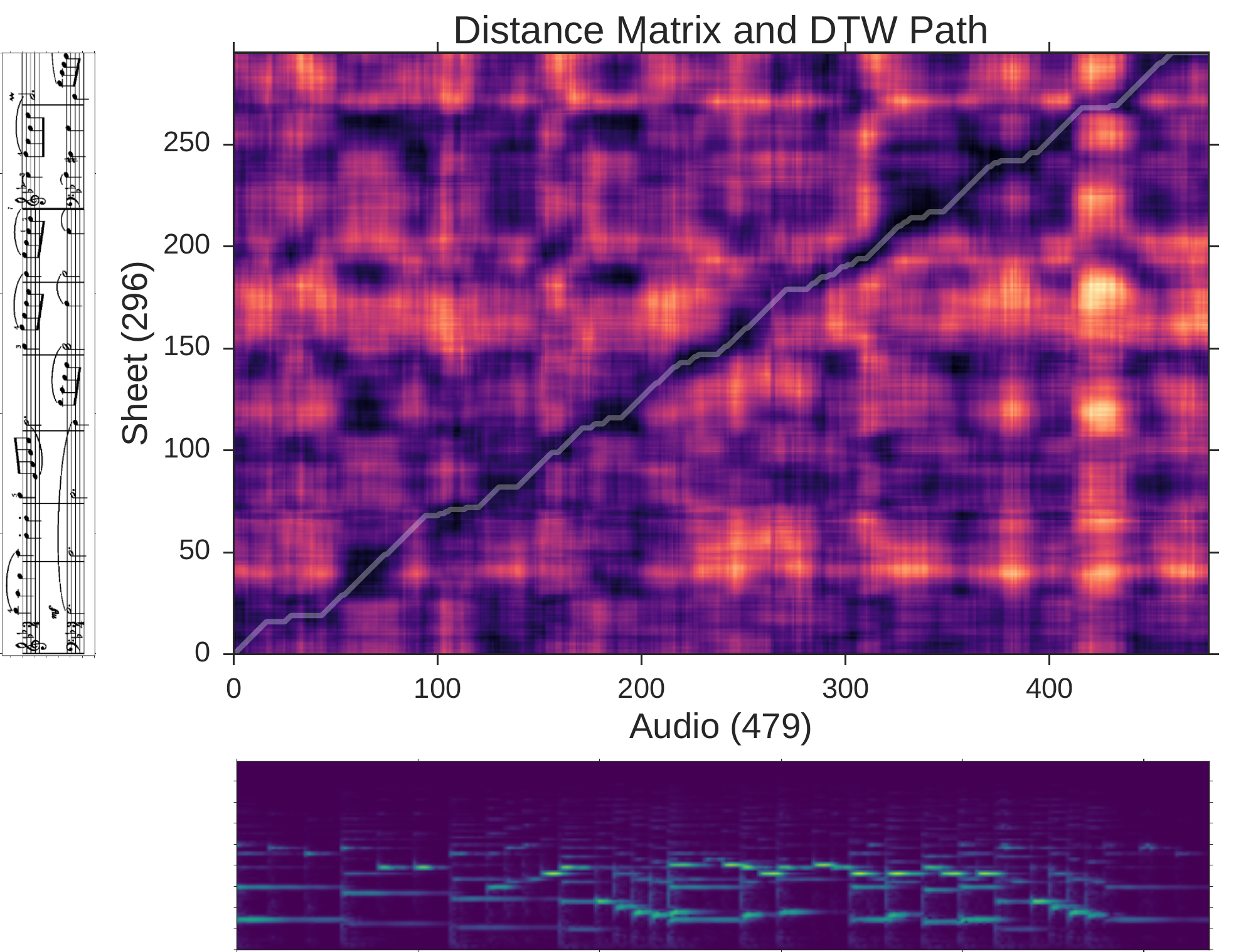}}
 \caption{Sketch of audio-to-sheet-music alignment by DTW on a similarity matrix
          computed on the embedding representation learned by the multi-modal matching network.
          The white line highlights the path of minimum costs through the sheet music given the audio.
}
\label{fig:dtw_path}
\end{figure}

For our task the two sequences to be aligned are the sequence of snippets from the sheet music image and the sequence of audio (spectrogram) excerpts, as described in Section \ref{subsec:data}.
The neural network presented in Section \ref{sec:methods} is then used to derive a local cost measure by computing the pairwise cosine distances between the embedded sheet snippets and audio excerpts (see Equation \ref{eq:cos_dist}).
The resulting cost matrix that relates all points of both sequences to each other is shown in Figure \ref{fig:dtw_path}, for a short excerpt from a simple Bach minuet. Then, the standard DTW algorithm is used to obtain the optimal alignment path.

\subsection{Evaluation of Approach}
For the evaluation we rely on the same dataset and setup as described above: learning the embedding only on Mozart, then aligning test pieces by Bach, Haydn, Beethoven, Chopin.
As evaluation measure we compute the absolute \emph{alignment error} (distance in pixels) of the estimated alignment to its ground truth alignment for each of the sliding window positions.
We further normalize the errors by dividing them by the sheet image width to be independent of image resolution.
As a naive baseline we compute a linear interpolation alignment which would correspond to a straight line diagonal in the distance matrix in Figure \ref{fig:dtw_path}.
We consider this as a valid reference as we do not consider repetitions for our experiments, yet (in which case things would become somewhat more complicated).
We further emphasize that the purpose of this experiment is to provide a proof of concept for this class of models in the context of sheet music alignment tasks, not to compete with existing specialized algorithms for music alignment.

\begin{figure}[t!]
 \centerline{\includegraphics[width=1.0\columnwidth]{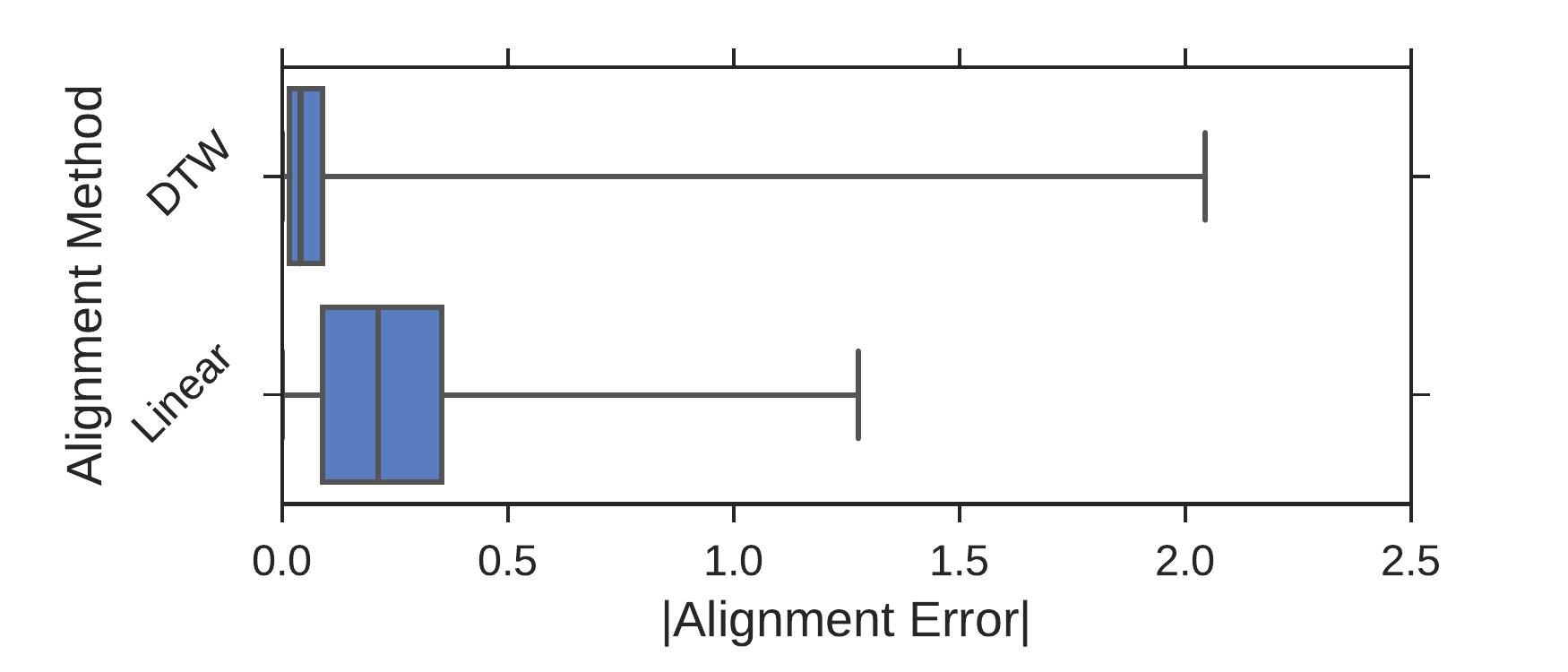}}
 \caption{Absolute alignment errors normalized by the sheet image width.
          We compare the linear baseline with a DTW on the cross-modality distance matrix
          computed on the embedded audio snippets and spectrogram excerpts.}
\label{fig:alignment_error}
\end{figure}

The results are summarized by the boxplots in Figure \ref{fig:alignment_error}.
The median alignment error for the linear baseline is 0.213 normalized image widths ($\approx 45$ mm in a printed page of sheet music).
When computing a DTW path through the distance matrix inferred by our mutimodal audio-sheet-music network this error decreases to 0.041 ($\approx$ 9 mm).
Note that values above 1.0 normalized page widths are possible as we handle a piece of sheet music as one single unrolled (concatenated) staff.

\section{Discussion and Conclusion}
\label{sec:conclusion}
We have presented a method for matching short excerpts of audio to their respective counterparts in sheet music images, via a multi-modal neural network that learns relationships between the two modalities,
and have shown 
how to utilize it for two MIR tasks: score identification from audio queries and offline audio-to-sheet-music alignment.
Our results provide a proof of concept for the proposed learning-retrieval paradigm
and lead to the following conclusions:
First, even though little training data is available, it is still possible to use
powerful state of the art image and audio models by designing appropriate (task specific) data augmentation strategies.
Second, as the best regularizer in machine learning is still a large amount of training data,
our results strongly suggest that annotating a truly large dataset will allow us to train general audio-sheet-music-matching models.
Recall that for this study we trained on only 13 Mozart pieces, and our model already started to generalize to unseen scores by other composers.

Another aspect of our method is that it works by projecting observations from different modalities into a very low-dimensional joint embedding space.
This compact representation is of particular relevance for the task of piece identification as our scoring function -- the cosine distance -- is a \emph{metric} that permits efficient search in large reference databases \cite{van2012metric}.
This identification-by-retrieval approach permits us to circumvent solving a large number of local DTW problems for piece identification as done, e.g., in \cite{fremerey2009sheet}.

For now, we have demonstrated the approach on sheet music of realistic complexity, but with synthesized audio (this was necessary to establish the ground truth).
The next challenge will be to deal with real audio and real performances, with challenges such as asynchronous onsets, pedal, and varying dynamics.

Finally, we want to stress that our claim is by no means that our proposal in its current stage is competitive with engineered approaches \cite{kurth2007automated,fremerey2009sheet,izmirli2012bridging} or methods relying on symbolic music or reference performances.
These methods have already proven to be useful in real world scenarios, with real performances \cite{arzt2015artificial}.
However, considering the progress that has been made in terms of score complexity (compared for example to the simple monophonic music used in \cite{Dorfer2016Towards}) we believe it is a promising line of research.

\section{Acknowledgements}
This work is supported by the Austrian Ministries BMVIT and BMWFW,
and the Province of Upper Austria via the COMET Center SCCH,
and by the European Research Council (ERC Grant Agreement
670035, project CON ESPRESSIONE).
The Tesla K40 used for this research was donated by the NVIDIA
corporation.

\bibliography{ISMIR2017_Audio2Score}

%
%
%
%

\end{document}